\documentclass[prd,twocolumn]{revtex4}
\usepackage{amsfonts}
\usepackage[mathscr]{eucal}

\begin{document}

\title{Purely twistorial string with canonical twistor field quantization}

\author{Sergey  Fedoruk$\,{}^{1}$\footnote{fedoruk@theor.jinr.ru},
Jerzy Lukierski$\,{}^{2}$\footnote{lukier@ift.uni.wroc.pl}}

\affiliation{\vspace{0.5cm} ${}^1$Bogoliubov Laboratory of  Theoretical Physics, JINR,
141980 Dubna, Moscow Region, Russia \\
${}^2$Institute for Theoretical Physics, University of Wroc{\l}aw, pl. Maxa Borna 9, 50-204
Wroc{\l}aw, Poland}

\begin{abstract}
\vspace{0.5cm} \noindent We introduce new purely twistorial scale--invariant action
describing the composite bosonic $D=4$ Nambu--Goto string with target space parametrized by
the pair of $D=4$ twistors. We show that by suitable gauge fixing of local scaling one gets
the bilinear twistorial action and canonical quantization rules for the two--dimensional
twistor--string fields. We consider the Poisson brackets of all constraints characterizing
our model and we obtain four first class constraints describing two Virasoro constraints
and two $U(1)\bigotimes U(1)$ Kac--Moody (KM) local phase transformations.

\end{abstract}
\vspace{2cm}

\maketitle

\setcounter{footnote}{0} \setcounter{equation}{0}
\section{Introduction}

The idea, due to Penrose (see e.g.~\cite{PenMac}), of using twistors as replacing the
primary space--time geometry, has been applied to many problems in physics, in particular
to the description of particles and
superparticles~\cite{Fer,Shir,BC,STV,Town,Berk,BetAz,Az,FFLM} as well as strings and
superstrings~\cite{Ceder,SSTV,BZ,Bars,Wit,Ber,BerWit,S,BAM}. For particle and superparticle
models the equivalence of twistorial and space--time approach has been shown on classical
and quantum level; for strings and superstrings the equivalence of two geometric pictures
was demonstrated only by comparing the spectra of quantized excitations (see
e.g.~\cite{Wit,Ber,BerWit}), without providing the equality of classical actions. The main
results following from the models of open tensionless twistor strings presented by
Witten~\cite{Wit} and Berkovits~\cite{Ber} was the close links with super-Yang-Mills gauge
theories, leading to new twistor representation of multihelicity tree and one-loop
amplitudes (see e.g.~\cite{CSW}). The tensionless closed superstring was also applied to
the description of the conformal supergravity sector~\cite{BerWit}; the relation with
standard (Poincare) supergravity amplitudes was subsequently obtained by adding the terms
breaking the conformal symmetry and introducing dimensionful parameter (Plank
mass)~\cite{AHM}. We should add also that the link between the Yang-Mills theory and
twistor string framework has been further advanced by providing the Yang-Mills action (in
$D=4$) in terms of fields on twistor space~\cite{M,BMS}.

It should be stressed that recent twistor string models developed
in~\cite{Wit,Ber,BerWit,S,BAM} were described by tensionless strings, which are the
one-dimensional counterpart of massless point particles. It is known that the description
of massive (super)particles requires in the Penrose framework the introduction of
two-(super)twistor geometry (see~\cite{BetAz,Az,AzLu}). Following~\cite{Ceder}, using
two-twistor target space purely twistorial tensionful string action was given~\cite{FL},
which is classically equivalent to Nambu--Goto (NG) string action with composite
space--time string fields. Unfortunately, our twistorial action from~\cite{FL} is
fourlinear, what presents a serious difficulty in performing the quantization procedure. In
this paper we resolve this difficulty. Following the reduction by reparametrization gauge
fixing of NG string to free string model, we show that the quadratic free twistor string
can be obtained by suitable gauge fixing of a new nonlinear twistor--string model. We
derive therefore the bilinear twistor--string model which leads for $D=2$ twistor--string
fields to standard Penrose quantization rules~\cite{Ceder,Gun-Kall}.

In our recent paper~\cite{FL}  we obtained purely twistorial fourlinear classical $D=4$
string action, with target space described by the two--twistor space. The fundamental
string world--sheet fields are described by the following pair of $D=4$ twistors
($A=1,\ldots,4$; $i=1,2$; $m=1,2$; $\alpha,\dot\alpha=1,2$)
\begin{eqnarray}\label{Z-com}
Z_{Ai}(\xi)&=&\left(\lambda_{\alpha i}(\xi),\, \mu_i^{\dot\alpha}(\xi)\right),
\\
\bar Z^{Ai}(\xi)&=&\left(\bar\mu^{\alpha i}(\xi),\, -\bar\lambda_{\dot\alpha}^{i}(\xi)
\right),\quad\bar\lambda_{\dot\alpha}^{i}=(\overline{\lambda_{\alpha i}}) ,\,\,
\bar\mu^{\alpha i}=(\overline{\mu_i^{\dot\alpha}})\nonumber
\end{eqnarray}
where $\xi^m=(\xi^0,\xi^1)$ denote the world--sheet coordinates; the indices $\alpha$,
$\dot\alpha$ and $i$ are lifted by $(2\times 2)$ antisymmetric tensor $\epsilon$,
$\epsilon^{12}=\epsilon_{21}=1$ ($\lambda^\alpha=\epsilon^{\alpha\beta}\lambda_\beta$,
$\lambda_\alpha=\epsilon_{\alpha\beta}\lambda^\beta$ etc.). In twistorial string model the
generalized Penrose relations~\cite{PenMac}, expressing the string phase space fields
$X_\mu(\xi)$, $P^m_\mu(\xi)$~\cite{Sieg,BH} ($\mu=1,\ldots,4$) by the twistor fields
$Z_{Ai}(\xi)$, $\bar Z^{Ai}(\xi)$ are the following
($P^m_{\alpha\dot\alpha}=\sigma^\mu_{\alpha\dot\alpha}P^m_\mu$)

\noindent $\bullet$ string extension of Cartan--Penrose formula
\begin{equation}\label{P-res-st}
P_{\alpha\dot\beta}^{\,m}(\xi)=e(\xi)\,
\tilde{\lambda}_{\dot\beta}^i(\xi)\,(\rho^m)_i{}^j\,\lambda_{\alpha j}(\xi).
\end{equation}
where $e=\det e_m^a$, ($e_m^a(\xi)$ is the zweibein), $(\rho^m)_i{}^j=e^m_a(\rho^a)_i{}^j$
($\rho^a$ are $2\times 2$ Dirac matrices; $a=1,2$), $\tilde{\lambda}_{\dot\alpha}^i =
\bar{\lambda}_{\dot\alpha}^j(\rho^0)_j{}^i$.

\noindent $\bullet$ string extension of Penrose incidence relations
\begin{equation}\label{Pen-inc}
\mu^{\dot\alpha}_i(\xi)=X^{\dot\alpha\beta}(\xi)\,\lambda_{\beta i}(\xi) ,\quad
\bar{\mu}^{\alpha i}(\xi)=\bar{\lambda}_{\dot\beta}^i(\xi)\,
X^{\dot\beta\alpha}(\xi).
\end{equation}

In~\cite{FL} it was shown that if we insert~(\ref{P-res-st}) into the first order phase
space formulation of Nambu--Goto string due to Siegel (see~\cite{Sieg})
\begin{equation}\label{act-s}
S=\!\int\!\! d^2\xi\left[P^{\,m}\partial_m X + {\textstyle
\frac{1}{2T}}(-h)^{-1/2}h_{mn}P^{\,m} P^{\,n}\right]
\end{equation}
one obtains the bosonic string model proposed by Soroka, Sorokin, Tkach and Volkov
(SSTV)~\cite{SSTV}
\begin{equation}\label{SSTV}
\!\!\! S\!=\!\int\!\! d^2\xi \, e\left[\tilde{\lambda}_{\dot\alpha} \rho^m\!\lambda_{\alpha
}\,
\partial_m X^{\dot\alpha\alpha}+ {\textstyle \frac{1}{2T}}\,(\lambda^{\alpha i}
\lambda_{\alpha i}) (\tilde{\lambda}_{\dot\alpha}^j\tilde{\lambda}^{\dot\alpha}_j)\right]
\end{equation}
where $T$ with mass dimension $[m^2]$ describes the string tension, occurring as proportionality factor in
Nambu--Goto string action.

We observe that the action~(\ref{SSTV}) is invariant under two local scale transformations
and two phase transformations:
\begin{equation}\label{loc-s-p-la}
\lambda^\prime_{\alpha k} =e^{c_k+i\varphi_k}\lambda_{\alpha k},\quad
\bar\lambda_{\dot\alpha}^{\prime\, k} =e^{c_k-i\varphi_k}\bar\lambda_{\dot\alpha}^k,
\end{equation}
\begin{equation}\label{zw-tr}
\begin{array}{rcl}
e^{\prime\, 1}_m +e^{\prime\, 2}_m &=&e^{-2c_1}(e^{1}_m +e^{2}_m), \\ \\
e^{\prime\, 1}_m
-e^{\prime\, 2}_m &=&e^{-2c_2}(e^{1}_m -e^{2}_m)
\end{array}
\end{equation}
where we use Weyl representation for $D=2$ Dirac matrices  $\rho_a$, with the products
$\rho_0\rho_a$ diagonal. In~\cite{FL} we suitably fixed one local scale transformation with
$c=c_1+c_2$ and one local phase transformation with $\varphi=\varphi_1+\varphi_2$ and
imposed the constraints
\begin{equation}\label{tw-norm}
\lambda^{\alpha i} \lambda_{\alpha i}=T= \bar\lambda_{\dot\alpha i}\bar\lambda^{\dot\alpha
i}.
\end{equation}
Subsequently after employing in~(\ref{SSTV}) the relations~(\ref{Pen-inc}) we obtained the
fourlinear purely twistorial Lagrangian~\cite{FL} which is induced on the world-sheet by
the following canonical 2-form
\begin{equation}\label{2-form}
\Theta^{(2)} =  \Theta^{(1)}_1 \wedge
\Theta^{(1)}_2
\end{equation}
where ($i=1,2$)~\cite{fut-1}
\begin{equation}\label{1-form}
\Theta^{(1)}_i = (\bar Z^{Ai} dZ_{Ai}- d\bar Z^{Ai} Z_{Ai})
\end{equation}
describes the standard twistorial one-form, defining the twistorial particle
actions~\cite{Shir,BetAz}.

In this paper we shall transform the action~(\ref{SSTV}) into 2--twistor action before any
gauge fixing. Further we shall fix suitably the local gauges~(\ref{loc-s-p-la}), in
different way than in~\cite{FL}. In such a way we shall cancel the fourlinear terms in the
action and we obtain the solvable gauge--fixed bilinear twistor--string action.

\setcounter{equation}{0}
\section{Purely twistorial picture of SSTV string model}

Let us apply the generalized incidence relations~(\ref{Pen-inc}) in order to remove
from~(\ref{SSTV}) the string space--time field $X_\mu(\xi)$. It follows
from~(\ref{Pen-inc}) that the twistor fields $Z_{Ai}(\xi)$, $\bar Z^{Ai}(\xi)$ should
satisfy four constraints ($(\overline{V_i^j})=-V_j^i$)
\begin{equation}\label{V-Z}
V_i^j \equiv Z_{Ai}\bar Z^{Aj}=\lambda_{\alpha i}\bar{\mu}^{\alpha j} - \mu^{\dot\alpha}_i
\bar{\lambda}_{\dot\alpha}^j \approx 0.
\end{equation}
Further, taking into account the incidence relations~(\ref{Pen-inc}) we obtain that
\begin{equation}\label{PX-Z}
\!\!\!\!\tilde{\lambda}_{\dot\alpha} \rho^m\!\lambda_{\alpha }\,
\partial_m X^{\dot\alpha\alpha}\! ={\textstyle\frac{1}{2}}
(\partial_m \tilde{Z}^{A}\rho^mZ_{A} -\tilde{Z}^{A}\rho^m\partial_m Z_{A} ).
\end{equation}
Inserting~(\ref{PX-Z}) into (\ref{SSTV}) and taking into consideration the
constraints~(\ref{V-Z}) we obtain the action
\begin{eqnarray}
 S \!\!&=&\!\! {\textstyle\frac{1}{2}}\int d\,^2\xi\,
e\left(\partial_m \tilde{Z}^{A}\rho^mZ_{A} -\tilde{Z}^{A}\rho^m\partial_m Z_{A} \right)- \,\,\nonumber\\
&& -\int d\,^2\xi\,\left( e{\textstyle\frac{M\bar M}{T}}+\Lambda_j^i V_i^j
\right)\label{act-zw}
\end{eqnarray}
where
\begin{equation}\label{A-Z-comp}
\begin{array}{rcl}
\phantom{\Big(} M&\equiv &\epsilon^{ij}I^{AB} Z_{Ai}Z_{Bj}=\lambda^{\alpha
i}\lambda_{\alpha i},
\\
\phantom{\Big(}\bar M &\equiv &-\epsilon_{ij}I_{AB}\bar Z^{Ai}\bar
Z^{Bj}=\bar\lambda_{\dot\alpha i}\bar\lambda^{\dot\alpha i}
\end{array}
\end{equation}
and $I^{AB}$ and $I_{AB}$ are the asymptotic twistors~\cite{PenMac}, described by singular
$4\times 4$ matrices $ I^{AB}=\left(
\begin{array}{cc}
  \epsilon^{\alpha\beta} & 0 \\
  0 & 0
\end{array}
\right)$, $I_{AB}=\left(
\begin{array}{cc}
  0 & 0 \\
  0 & \epsilon^{\dot\alpha\dot\beta}
\end{array}
\right)\,. $

Let us indicate here that Berkovits~\cite{Ber} and Siegel twistor string~\cite{S} models
can be obtained from the action~(\ref{act-zw}) in particular tensionless limit (see
also~\cite{BAM}). In Weyl representation for $d=2$ gamma-matrices the action can be
rewritten in the form
\begin{eqnarray}
S \!\!&=&\!\! \int d\,^2\xi\,
ee^m_{-}\left(\bar{Z}^{A1} \partial_m Z_{A1} -\partial_m \bar{Z}^{A1} Z_{A1} \right)+ \,\,\nonumber\\
&&\!\! +\int d\,^2\xi\,
ee^m_{+}\left(\bar{Z}^{A2} \partial_m Z_{A2} - \partial_m \bar{Z}^{A2} Z_{A2} \right)- \,\,\nonumber\\
&& -\int d\,^2\xi\,\left( e{\textstyle\frac{M\bar M}{T}}+\Lambda_j^i V_i^j \right)
\label{act-zw-W}
\end{eqnarray}
where $e^m_{\pm}=\frac12(e^m_{0}\pm e^m_{1})$. Then, making rescaling $Z_{Ai}\rightarrow
T^{1/2} Z_{Ai}$, $e^m_{-} \rightarrow T^{-1} e^m_{-}$, $e^m_{+} \rightarrow T e^m_{+}$,
$\Lambda_1^1 \rightarrow T^{-1} \Lambda_1^1$ and taking tensionless limit $T \rightarrow 0$
we obtain Siegel twistor string action~\cite{S,fut-2}
\begin{equation}\label{act-Sieg-tw}
S_{T\to 0} = \int d\,^2\xi\, e\left(\bar{Z}^{A1} \nabla_{\!-} Z_{A1} -\nabla_{\!-}
\bar{Z}^{A1} Z_{A1} \right)
\end{equation}
where $\nabla_{\!-}=e^m_{-} \partial_m +iA$ is the world-sheet covariant derivative with
$U(1)$--connection $A=\bar A=\frac{i}{2e}\Lambda_1^1$. One can add that by putting
in~(\ref{act-zw-W}) $Z_{A2}= 0$ one obtains as well the action~(\ref{act-Sieg-tw}). The
Berkovits model~\cite{Ber} can be considered as double variant of the Siegel model~\cite{S}
with summ of two actions~(\ref{act-Sieg-tw}), one for left moving, and second for right
moving twistor string fields~(\ref{Z-com}).

The variation of~(\ref{act-zw}) with respect to the zweibein $e_m^a$ gives the relation
\begin{equation} \label{e-resol}
e_m^a = {\textstyle\frac{T}{M\bar M}} \left(\tilde{Z}^{A}\rho^a\partial_m Z_{A} -\partial_m
\tilde{Z}^{A}\rho^aZ_{A} \right).
\end{equation}
Inserting~(\ref{e-resol}) into (\ref{act-zw}) we obtain our new nonlinear twistor string
model
\begin{eqnarray}\label{act}
S &=& \int d\,^2\xi\,\Big[{\textstyle\frac{T}{M\bar M}}\,\epsilon^{mn} \left(\partial_m
Z_{A1}\bar Z^{A1} - Z_{A1}\partial_m\bar Z^{A1} \right)\times \nonumber \\
&& \times\left(\partial_n Z_{B2} \bar Z^{B2}  - Z_{B2}\partial_n\bar Z^{B2}\right)
-\Lambda_j^i V_i^j \Big],
\end{eqnarray}
where $\Lambda_i^j=-(\overline{\Lambda_j^i})$.

It should be observed that the twistor string fields have different mass dimensions in
comparison with the twistor coordinates in particle mechanics. For consistency we should
assume in~(\ref{P-res-st}), (\ref{Pen-inc}) that
$$
[\lambda_{\alpha i}]=m^1,\qquad [\mu_i^{\dot\alpha}]=m^0.
$$

Lagrangian density in the action~(\ref{act}) can be represented in the following form (we use the
usual notations: $\dot X \equiv \frac{\partial X}{\partial \tau}$,
$\acute{X} \equiv \frac{\partial X}{\partial \sigma}$ where $\tau\equiv\xi^0$, $\sigma\equiv\xi^1$)
\begin{equation}\label{L0-1}
\begin{array}{rcl}
{\cal L}&=&{\textstyle \frac{T}{M\bar M}} \, Q_2  \left( \dot Z_{1} \bar Z^{1} - Z_{1}
\dot{\bar Z}{}^{1} \right) -
\\
&&-{\textstyle \frac{T}{M\bar M}} \, Q_1 \left( \dot Z_{2} \bar Z^{2} - Z_{2} \dot{\bar
Z}{}^{2} \right) -\Lambda_j^i V_i^j
\end{array}
\end{equation}
where
$$
Q_1 \equiv \acute{Z}_{1} \bar Z^{1} - Z_{1} \acute{\bar Z}{}^{1}, \qquad Q_2 \equiv
\acute{Z}_{2} \bar Z^{2} - Z_{2} \acute{\bar Z}{}^{2}
$$
(we omit repeated index
$A$ in these expressions and below). We see that the definitions of the momenta $ P^{Ai} =
{\partial {\cal L}}/{\partial\dot Z_{Ai}}$, $\bar P_{Ai} = {\partial {\cal
L}}/{\partial\dot{\bar Z}{}^{Ai}} $ introduce the constraints
\begin{equation}\label{D-com-nf}
\begin{array}{rcl}
\mathscr{D}^{A1} &\equiv& P^{A1} - {\textstyle \frac{T}{M\bar M}} \, Q_2 {\bar Z}^{A1}
 \approx 0, \\
 &&\\
  \mathscr{D}^{A2} &\equiv& P^{A2} + {\textstyle
\frac{T}{M\bar M}} \, Q_1 {\bar Z}^{A2}  \approx 0,
\\
&&\\
\bar\mathscr{D}_{A1} &\equiv& \bar P_{A1} + {\textstyle \frac{T}{M\bar M}} \, Q_2 Z_{A1}
\approx 0, \\
&&\\
 \bar\mathscr{D}_{A2} &\equiv& \bar P_{A2} - {\textstyle \frac{T}{M\bar
M}} \, Q_1 Z_{A2} \approx 0.
\end{array}
\end{equation}

The constraints~(\ref{V-Z}) and~(\ref{D-com-nf}) describe the set of primary constraints of
the model~(\ref{L0-1}). From~(\ref{D-com-nf}) we get the following two first class
constraints
\begin{equation}\label{1-dil}
\begin{array}{rcl}
\!\!\!\!\!\mathscr{F}_{1}\!&=&\!Z_{A1}\mathscr{D}^{A1} + {\bar Z}^{A1}\bar\mathscr{D}_{A1}=
Z_{A1}P^{A1} + {\bar Z}^{A1}\bar P_{A1},
\\
&& \\
 \!\!\!\!\!\mathscr{F}_{2}\!&=&\!Z_{A2}\mathscr{D}^{A2} + {\bar Z}^{A2}\bar\mathscr{D}_{A2}=
Z_{A2}P^{A2} + {\bar Z}^{A2}\bar P_{A2}
\end{array}
\end{equation}
which generate the local scale transformations (see also~(\ref{V-Z})) \cite{fut-1}
\begin{equation}\label{sc-Z}
\delta Z_{Ai} =c_i Z_{Ai} ,\qquad \delta \bar Z^{Ai} = c_i \bar Z^{Ai}.
\end{equation}

We observe that the nonlinearity in~(\ref{D-com-nf}) is not invariant under the two scale
gauge transformations~(\ref{sc-Z}). One gets
\begin{equation}\label{sc-nl}
\begin{array}{rcl}
\delta\left({Q_1}/{M\bar M}\right)&=& -2 c_2\left({Q_1}/{M\bar M}\right),
\\ \delta\left({Q_2}/{M\bar M}\right)&=& -2 c_1\left({Q_2}/{M\bar M}\right).
\end{array}
\end{equation}
If we supplement~(\ref{sc-Z}) by suitable variation of Lagrange multipliers ($ \delta
\Lambda_i^j =-(c_i+c_j) \Lambda_i^j $) we obtain that the complete twistor--string
Lagrangian~(\ref{L0-1}) is invariant under the local scaling described by~(\ref{sc-Z}).

\setcounter{equation}{0}
\section{Scale gauge fixing and canonical quantization} Let us put the following
gauge fixing conditions for the gauge transformations~(\ref{sc-Z}), (\ref{sc-nl})
\begin{equation}\label{ga-con}
(TQ_1)/(M\bar M)= -1/2,\qquad (TQ_2)/(M\bar M)=1/2.
\end{equation}

As a result, in this gauge we obtain from~(\ref{L0-1}) the bilinear gauge--fixed
twistor--string action $S_{gf}$:
\begin{equation}\label{act-fir}
S_{gf} = \int d\,^2\xi\,\left[ {\textstyle \frac{1}{2}}\left(\dot Z_{i} \bar Z^{i} - Z_{i}
\dot{\bar Z}{}^{i}\right) -\Lambda_j^i V_i^j -\Lambda_i \Phi_i\right]
\end{equation}
where $Z_{Ai} $, $\bar Z^{Ai}$ are restricted by the constraints~(\ref{V-Z}) and the gauge
fixing~(\ref{ga-con})
\begin{equation}\label{Q1,2-con+}
\begin{array}{rcl}
\Phi_1 &=& {\textstyle \frac{1}{2}}\,( \acute{Z}_{1} \bar Z^{1} - Z_{1}
\acute{\bar Z}{}^{1} ) + {\textstyle \frac{M\bar M}{4T}} \approx 0, \\
&& \\
\Phi_2&=& {\textstyle \frac{1}{2}}\,( \acute{Z}_{2} \bar Z^{2} - Z_{2} \acute{\bar Z}{}^{2}
)-{\textstyle \frac{M\bar M}{4T}} \approx 0 .
\end{array}
\end{equation}
Let us observe that similarly like the mass parameter in phase space formulation of
relativistic particle action, the tension parameter $T$ enters into the constraints
(see~(\ref{Q1,2-con+})).

In the gauge~(\ref{ga-con}) the constraints~(\ref{D-com-nf}) have the form
\begin{equation}\label{D-com}
\begin{array}{rcl}
\mathscr{D}^{Ai} &\rightarrow & D^{Ai} \equiv P^{Ai} - {\textstyle \frac{1}{2}}\,{\bar
Z}^{Ai}
 \approx 0, \\
 &&\\
\bar\mathscr{D}_{Ai} &\rightarrow & \bar D_{Ai} \equiv \bar P_{Ai} + {\textstyle
\frac{1}{2}}\,Z_{Ai} \approx 0 .
\end{array}
\end{equation}

If we use canonical equal time Poisson brackets
\begin{eqnarray}
 \{ Z_{Ai}(\sigma),
P^{Bj}(\sigma^\prime)\}_{{}_{\rm P}}&=& \delta_A^B \delta_i^j \delta(\sigma -
\sigma^\prime), \nonumber
\\
\{{\bar Z}^{Ai}(\sigma), \bar P_{Bj}(\sigma^\prime) \}_{{}_{\rm P}}&=& \delta^A_B
\delta^i_j \delta(\sigma - \sigma^\prime) \nonumber
\end{eqnarray}
we obtain the brackets
\begin{equation}\label{D-bD}
\{ D^{Ai}(\sigma), \bar D_{Bk}(\sigma^\prime) \}_{{}_{\rm P}}=-\delta_B^A
\delta^k_i\delta(\sigma - \sigma^\prime).
\end{equation}
Inserting into Dirac brackets (we denote it by ``T'' as twistor brackets):
\begin{eqnarray}
&&\{ A(\sigma), B(\sigma^\prime) \}_{{}_{\rm T}}= \{ A(\sigma), B(\sigma^\prime)
\}_{{}_{\rm P}}- \label{Dir}\\
&&\quad-\int d\sigma_1 \{ A(\sigma), D^{Ai}(\sigma_1) \}_{{}_{\rm P}}
\{ \bar D_{Ai}(\sigma_1), B(\sigma^\prime) \}_{{}_{\rm P}}+ \nonumber\\
&&\quad +\int d\sigma_1 \{ A(\sigma), \bar D_{Ai}(\sigma_1) \}_{{}_{\rm P}} \{
D^{Ai}(\sigma_1), B(\sigma^\prime) \}_{{}_{\rm P}} \nonumber
\end{eqnarray}
we arrive at the {\it standard twistor canonical relations} for free twistorial string
\begin{equation}\label{TCB}
\{ Z_{Ai}(\sigma), {\bar Z}^{Bj}(\sigma^\prime) \}_{{}_{\rm T}}= \delta_A^B \delta_i^j\,
\delta(\sigma - \sigma^\prime)
\end{equation}
which were assumed e.g. a priori in~\cite{Ceder}, but not derived from the twistorial
string action.

In such a way we obtained free two dimensional twistor model which corresponds e.g. to the
twistor string formulation given in~\cite{Gun-Kall}. It should be added that in the free
twistor string action~(\ref{act-fir}) the constraints are derived, in similar way as
Virasoro conditions in Nambu-Goto string framework, by gauge-fixing procedure which leads
uniquely to bilinear action.

Using the relations~(\ref{TCB}) one can calculate the PB of primary
constraints~(\ref{V-Z}), (\ref{Q1,2-con+}). One obtains
\begin{equation}\label{V-V}
\{ V_i^j(\sigma), V_k^l(\sigma^\prime) \}_{{}_{\rm T}}= -\left(\delta_k^j\,
V_i^l-\delta_i^l\, V_k^j\right)\delta(\sigma - \sigma^\prime)\,,
\end{equation}
\begin{equation}
\!\!\begin{array}{rcl} \{ \Phi_+(\sigma), \Phi_+(\sigma^\prime) \}_{{}_{\rm
T}}\!\!\!\!&=&\!\!\! \Big( \Phi_+(\sigma)+ \Phi_+(\sigma^\prime)\Big)\delta^\prime(\sigma -
\sigma^\prime)\,,
\\
\{ \Phi_+(\sigma), \Phi_-(\sigma^\prime) \}_{{}_{\rm T}}\!\!\!\!&=&\!\!\! \Big(
\Phi_-(\sigma)+ \Phi_-(\sigma^\prime)\Big)\delta^\prime(\sigma - \sigma^\prime)\,,
\\
\{ \Phi_-(\sigma), \Phi_-(\sigma^\prime) \}_{{}_{\rm T}}\!\!\!\!&=&\!\!\! \Big(
\Phi_+(\sigma)+ \Phi_+(\sigma^\prime)\Big)\delta^\prime(\sigma - \sigma^\prime)\,,
\end{array}\!\!\!
\end{equation}
\begin{equation}
\{ \Phi_+(\sigma), V_i^j(\sigma^\prime) \}_{{}_{\rm T}}=V_i^j(\sigma)\delta^\prime(\sigma -
\sigma^\prime)\,,
\end{equation}
\begin{equation}\label{Phi--V}
\begin{array}{lcl}
\phantom{\Big(}\{ \Phi_-(\sigma), V_1^1(\sigma^\prime) \}_{{}_{\rm
T}}&=&V_1^1(\sigma)\delta^\prime(\sigma - \sigma^\prime)\,, \\
\phantom{\Big(} \{ \Phi_-(\sigma), V_2^2(\sigma^\prime) \}_{{}_{\rm
T}}&=&-V_2^2(\sigma)\delta^\prime(\sigma - \sigma^\prime)\,,
\\ \phantom{\Big(}
\{ \Phi_-(\sigma) , V_1^2(\sigma^\prime) \}_{{}_{\rm T}}&=& -Q_1^2(\sigma)\delta(\sigma -
\sigma^\prime) \,, \\ \phantom{\Big(}
 \{ \Phi_- (\sigma) , V_2^1(\sigma^\prime) \}_{{}_{\rm T}}&=&
Q_2^1(\sigma)\delta(\sigma - \sigma^\prime)
\end{array}
\end{equation}
where $\Phi_{\pm}=\Phi_1\pm \Phi_2$ and
\begin{equation}\label{def-Q12}
\begin{array}{lcl}
\phantom{\Big(} Q_1^2 &\equiv& \acute Z_{A1}\bar Z^{A2} -  Z_{A1}\acute{\bar Z}{}^{A2}\,, \\
\phantom{\Big(} Q_2^1 &\equiv& \acute Z_{A2}\bar Z^{A1} -  Z_{A2}\acute{\bar Z}{}^{A1}\,.
\end{array}
\end{equation}

Interestingly enough we see that the constraints~(\ref{Q1,2-con+}) describe Virasoro
algebra, and the constraints $V_i^j$~(\ref{V-Z}) form the $U(2)$ Kac--Moody algebra. The
cross relations between these algebras are however not closed due to the last two PB in
relations~(\ref{Phi--V}) where do appear the bilinears $Q_1^2$, $Q_2^1$. We shall show in
Sect. IV that the bilinears~(\ref{def-Q12}) define secondary constraints in our model.

\setcounter{equation}{0}
\section{Primary and secondary constraints}

Further we shall consider Hamiltonian formulation of our twistorial action~(\ref{act}) in
the gauge~(\ref{ga-con}), i.e. with the constraints $\mathscr{D}^{Ai}$,
$\bar\mathscr{D}_{Ai}$ replaced by $D^{Ai}$, $\bar D_{Ai}$. Remaining primary constraints
$\Phi_i$, $V_i^j$ are described by the relations~(\ref{V-Z}), (\ref{Q1,2-con+}). The
Hamiltonian corresponding to the action~(\ref{act-fir}) looks as follows
\begin{equation}\label{H1}
H_1= \int d\sigma \left( \Lambda_{Ai} D^{Ai}+ \bar\Lambda^{Ai}\bar D_{Ai} +\Lambda_j^i
V_i^j + \Lambda^{\!i} \Phi_i \right).
\end{equation}
Because the twistor--string momenta $P^{Ai}$, $\bar P_{Ai}$ are entering only into
constraints~(\ref{D-com}), the nonvanishing canonical PB of the constraints are only those
with $D^{Ai}$, $\bar D_{Ai}$.

The preservation of the constraints $D^{Ai}$, $\bar D_{Ai}$ in time ($\dot D^{Ai}=\left\{
D^{Ai}, H_1 \right\}_{{}_{\rm P}}\approx 0$, $\dot {\bar D}_{Ai}=\left\{ {\bar D}_{Ai}, H_1
\right\}_{{}_{\rm P}}\approx 0$) leads to the expressions for $\Lambda_{Ai}$,
$\bar\Lambda^{Ai}$ as suitable linear combinations of the Lagrange multipliers
$\Lambda_j^i$, $\Lambda^i$. The time independence of the remaining constraints~(\ref{V-Z}),
(\ref{Q1,2-con+}) ($\dot F_M=\left\{ F_M, H_1 \right\}_{{}_{\rm P}} \approx 0$; $
F_M=(\Phi_i,V_i^j)$) leads after long but simple algebraic calculation to the following
conditions ($i=1,2$)
\begin{eqnarray}
\dot \Phi_i \approx 0 &\Rightarrow&
\Lambda_1^2 Q_2^1 - \Lambda_2^1 Q_1^2 \approx 0, \label{co-Phi}\\
&& \nonumber\\
\dot V_i^j \approx 0 &\Rightarrow& \Lambda^{\!\!-}\, Q_i^j \approx 0, \quad i\neq
j\label{co-V12}. \label{co-V21}
\end{eqnarray}
where $\Lambda^-\equiv\Lambda^1-\Lambda^2$. Vanishing of time derivatives of the
constraints $V_1^1$, $V_2^2$ do not require additional relations.

After substitution of $\Lambda_{Ai}$, $\bar\Lambda^{Ai}$ in terms of remaining Lagrange
multipliers the Hamiltonian~(\ref{H1}) takes the form
\begin{equation}\label{H1a}
H_1= \int d\sigma \left( \Lambda_j^i \hat V_i^j + \Lambda^{\!i} \hat\Phi_i \right)
\end{equation}
where
\begin{equation}\label{hat-V}
\hat V_{i}^j \equiv  Z_i P^j - \bar P_i \bar Z^j
 ,
\end{equation}
\begin{equation}\label{hat-Phi-}
\begin{array}{rcl}
\hat\Phi_1 &\equiv & {\textstyle\frac{1}{2}}\,( \acute{Z}_1 P^1 - Z_1
\acute{P}^1 + \bar P_1 \acute{\bar Z}{}^1 - \acute{\bar P}_1 \bar Z^1) + {\mathcal R},\\
&&\\
\hat\Phi_2 &\equiv & {\textstyle\frac{1}{2}}\,( \acute{Z}_2 P^2 - Z_2 \acute{P}^2 + \bar
P_2 \acute{\bar Z}{}^2 - \acute{\bar P}_2 \bar Z^2) - {\mathcal R},
\end{array}
\end{equation}
$$
{\mathcal R}\equiv {\textstyle\frac{M}{4T}} (P^1I\bar Z^2 +\bar Z^1 I P^2)-
{\textstyle\frac{\bar M}{4T}} (\bar P_1I Z_2 + Z_1 I \bar P_2) - {\textstyle\frac{M\bar
M}{4T}} .
$$
One can check that the constraints $\hat\Phi_i$, $\hat V_{i}^j$ differ from the constraints
$\Phi_i$, $V_{i}^j$ by terms linear in constraints $D^{Ai}$, $\bar D_{Ai}$.

The equations~(\ref{co-Phi})--(\ref{co-V21}) describe the additional restrictions. There
are two possible choices:

{\bf i)}
We choose $\Lambda^- \neq 0$ and the secondary constraints
\begin{equation}\label{con-Q12}
Q_1^2\approx 0,\qquad Q_2^1\approx 0 .
\end{equation}
In such a case one should add to~(\ref{H1})  the secondary constraints $Q_1^2\approx 0$,
$Q_2^1\approx 0$, and check the closure for arbitrary time.

{\bf ii)} One can choose alternatively $Q_1^2=(\overline{Q_2^1})\neq 0$ and
$$
\Lambda^-= 0\,,\qquad \Lambda_1^2-\Lambda_2^1= (\Lambda_1^2+\Lambda_2^1) (Q_1^2-Q_2^1)/
(Q_1^2+Q_2^1).
$$
In such a case the closure of the constraints algebra at arbitrary time implies the change
of the nature of two primary constraints from first class to second class. One can show
that in such a case the number of degrees of freedom of tensor string will not coincide
with the number of physical degrees of freedom of bosonic string. Further we shall study
only the case~{i)}.

In order to show that after adding~(\ref{con-Q12}) we obtained complete set of constraints
we should consider the second stage Hamiltonian
\begin{equation}\label{H2}
H_2= H_1+\int d\sigma \left( L^1_2 Q_1^2 +L^2_1 Q_2^1\right)
\end{equation}
where $H_1$ is defined in~(\ref{H1}).

The preservation of the constraints $D^{Ai}$, $\bar D_{Ai}$ in time ( $\dot D^{Ai}=\left\{
D^{Ai},H_2\right\}_{{}_{\rm P}} \approx 0$, $\dot {\bar D}_{Ai}=\left\{ {\bar
D}_{Ai},H_2\right\}_{{}_{\rm P}} \approx 0$) leads again to the formulae, expressing
$\Lambda_{Ai}$, $\bar\Lambda^{Ai}$ by means of the linear combination of the Lagrange
multipliers $\Lambda_j^i$, $\Lambda^i$, $L^1_2$ and $L^2_1$. After substitution of these
formulae in~(\ref{H2}) we can check that the constraints $\hat V_{i}^j$, $\hat\Phi_i$
remain the same, i.e. are given by~(\ref{hat-V}), (\ref{hat-Phi-}), but the secondary
constraints~(\ref{con-Q12}) are modified and have the form
\begin{equation}\label{hat-Q12}
\begin{array}{rcl}
\hat Q_1^2 &\equiv & \acute{Z}_1 P^2 + \bar P_1 \acute{\bar Z}{}^2 - Z_1 \acute{P}^2
-\acute{\bar P}_1 \bar Z^2
,\\
&&\\
\hat Q_2^1 &\equiv & \acute{Z}_2 P^1 + \bar P_2 \acute{\bar Z}{}^1 - Z_2 \acute{P}^1
-\acute{\bar P}_2 \bar Z^1
 .
\end{array}
\end{equation}

Time independence of other constraints ($F_M=( V_i^j, \Phi_i, Q_1^2,Q_2^1)$; $\dot
F_M=\left\{F_M, H_2 \right\}_{{}_{\rm P}} \approx 0$) leads to the following new four
conditions:
\begin{eqnarray}
\dot V_i^j \approx 0 &\Rightarrow& M\bar M L_i^j \approx 0, \quad i\neq
j,\label{nco-V12}\\
&& \nonumber\\
\dot Q_i^j \approx 0 &\Rightarrow& M\bar M \Lambda_i^j \approx 0, \quad i\neq
j.\label{nco-Q12}
\end{eqnarray}

From~(\ref{nco-V12})--(\ref{nco-Q12}) follows the vanishing of $L_1^2$, $L_2^1$,
$\Lambda_1^2$ and $\Lambda_2^1$ in the Hamiltonian~(\ref{H2}). As a result, in comparison
with the Hamiltonian~(\ref{H1a}), the constraints $\hat V_1^2$, $\hat V_2^1$ change their
nature (from first to second class) and in final Hamiltonian~(\ref{H2}) remain only $\hat
V_1^1$, $\hat V_2^2$, $\hat \Phi_i$ as first class constraints. These four constraints have
the following canonical nonvanishing equal time PB brackets~\cite{fut-1}:
\begin{equation}
\Big\{ \hat \Phi_i(\sigma), \hat \Phi_j(\sigma^\prime) \Big\}_{{}_{\rm
P}}\!\!=\delta_{ij}\Big( \hat \Phi_i(\sigma)+ \hat
\Phi_i(\sigma^\prime)\Big)\delta^\prime(\sigma - \sigma^\prime),
\end{equation}
\begin{equation}
\Big\{ \hat \Phi_i(\sigma), \hat V_j^j(\sigma^\prime) \Big\}_{{}_{\rm
P}}\!\!=\delta_{ij}\hat V_i^i(\sigma)\delta^\prime(\sigma - \sigma^\prime).
\end{equation}
Calculating at a given time $\tau$ the Poisson brackets of $Z_{Ai}$, $\bar Z^{Ai}$ with the
four--parameter generator of local symmetry transformations in our model
\begin{equation}
\sum_{k=1}^2\int d\sigma \left( \varepsilon_k(\sigma,\tau)\hat \Phi_k(\sigma,\tau)+
i\varphi_k(\sigma,\tau)\hat V_k^k(\sigma,\tau)\right)
\end{equation}
one obtains using e.g. the considerations in~\cite{BH} (see Sect. 12.2.2), that the
functions $\varepsilon_i$ describe infinitesimal local world--sheet transformations, and
$\varphi_i$ lead to the Abelian phase transformations~(\ref{loc-s-p-la}).

Using formulae~(\ref{hat-V})--(\ref{hat-Phi-}) and (\ref{hat-Q12}) one can also calculate
the canonical PB matrix of second class constraints $D^{Ai}$, $\bar D_{Ai}$, $\hat V_1^2$,
$\hat V_2^1$ and $\hat Q_1^2$, $\hat Q_2^1$. If we observe that the canonical PB of the
constraints $D^{Ai}$, $\bar D_{Ai}$ with all other constraints $\hat \Phi_i$, $\hat V_i^j$,
$\hat Q_1^2$, $\hat Q_2^1$ are proportional to $D^{Ai}$, $\bar D_{Ai}$, we can show that
the Dirac bracket eliminating the constraints $D^{Ai}$, $\bar D_{Ai}$ (twistor brackets)
provide the same algebra of all the constraints as the canonical PB.

Let us calculate finally in our model the number of physical degrees of freedom. The
unconstrained phase space field variables $Z_{Ai}$, ${\bar Z}^{Ai}$ contains { sixteen}
field variables (after introducing twistor brackets which eliminate $P^{Ai}$, ${\bar
P}_{Ai}$). The second class constraints $\hat V_1^2$, $\hat V_2^1$, $\hat Q_1^2$, $\hat
Q_2^1$ remove four, and the first class constraints $\hat V_1^1$, $\hat V_2^2$, $\hat
\Phi_i$ remove { eight} degrees of freedom. In conclusion, we have { four} fields
describing physical real degrees of freedom, as in the case of Nambu--Goto string.

\setcounter{equation}{0}
\section{Final remarks}

One of important problems of twistorial formulation of string theory is its relation with
standard string theory. In this paper we propose to relate these two descriptions of
reparametrization--invariant two--dimensional elementary objects in very close way -- it
appears that already on classical level one can relate the twistor and space--time actions
by suitable nonlinear change of variables. In this paper we find new formulation of the
twistor string model, which permits to obtain by suitable gauge fixing the bilinear
twistorial Lagrangian and the standard twistorial commutation relations (see~(\ref{TCB})).
In such a way we proceed in analogous way as in Nambu--Goto formulation of string model in
order to get the solvable bilinear action.

In this paper we performed the complete constraint analysis of our twistor string model and
derived the set of local symmetries. After suitable gauge fixing and performing the
constraints analysis we are left with a pair of Virasoro algebras, describing the
reparametrization of the local world--sheet parameters; besides there are present two KM
generators $U(1)\bigotimes U(1)$, producing the change of local phases of two twistors.

There are several problems which could be further studied:

{\bf i)} In our case we consider the linear Lie--algebraic closure of four first class
constraints. It is interesting to study if there exists a closure of coupled two Virasoro
and four $U(2)$ KM generators $V_i^j$ (six constraints!) in the framework with nonlinearly
extended Poisson structures. In such framework one can look for the comparison with
interesting considerations in~\cite{Ceder}, where the twistorial string was constructed
without action, by postulating six constraints in the twistorial string phase space.

{\bf ii)} In~\cite{Gun-Kall} it was assumed that the twistorial model analogous
to~(\ref{act-fir}) can be useful for the description of string--like quarks, as
two--dimensional fundamental $SU(2,2|4)$ fields. It should be added that the twistor-string
models in~\cite{Wit,Ber,BerWit,S,BAM,CSW,AHM} are all supersymmetric in twistorial target
space, mostly with $N=4$ $D=4$ supersymmetry. Because of known theoretical advantages of
superstring models it is interesting to extend our scheme, in particular to $D=10$ $N=1$
superstring, and possibly perform the dimensional reduction $D=10 \rightarrow
D=4$~\cite{fut-3}.

{\bf iii)} In this paper we discuss the nonchiral bosonic strings. One can also develop our
scheme for the twistor strings constructed from two left- (right-) handed twistor fields
$Z_{Ai}(\xi_+)$ ($Z_{Ai}(\xi_-)$), where $\xi_\pm =\tau\pm\sigma$, and consider their
supersymmetric extensions. In such a way we would achieve closer link with the twistor
string models, proposed in~\cite{Ber,S}.

\begin{acknowledgements}
The authors would like to thank Martin Cederwall for interesting remarks. This paper has
been supported by Russian--Polish Bogoliubov--Infeld Programme, Polish Ministry Of Science
and Higher Education grant NN 202318534 (J.L.) and by RFBR grants 06-02-16684, 08-02-90490
and INTAS grant 05-79-28 (S.F.).
\end{acknowledgements}



\begin{thebibliography}{99}
\bibitem{fut-2}
Yang-Mills current part of Siegel action we do not considered here.
\bibitem{fut-1}
In these formulas we do not summ the repeated indices $i$, $j$.
\bibitem{fut-3}
Such considerations should be related with the approaches presented e.g. in~\cite{BAPV,Uv}.
\end{thebibliography}
\end{document}